\documentclass[10pt, conference, compsocconf]{IEEEtran}
\IEEEoverridecommandlockouts

\usepackage{cite}
\usepackage{amsmath,amssymb,amsfonts}
\usepackage{graphicx}
\usepackage{textcomp}
\usepackage{xcolor}
\usepackage{soul}
\usepackage{url}
\usepackage{fancyhdr, lipsum}
\usepackage{setspace}
\let\labelindent\relax
\usepackage{enumitem}
\usepackage[symbol]{footmisc}
\def\BibTeX{{\rm B\kern-.05em{\sc i\kern-.025em b}\kern-.08em
    T\kern-.1667em\lower.7ex\hbox{E}\kern-.125emX}}

\usepackage{verbatim}

\usepackage{tikz}
\usepackage{pgfplots}
\usepackage{algorithmicx} 
\usepackage[normalem]{ulem}
\usepackage{algcompatible}
\usepackage{algpseudocode}
\usepackage{xcolor}
\usepackage{threeparttable}
\usepackage{pifont}
\usepackage{etaremune}
\usepackage{caption}
\usepackage{subcaption}
\usepackage[export]{adjustbox}
\usepackage{listings}
\usepackage{courier}

\usepackage{algorithm} 
\usepackage{algorithmicx}
\usepackage{fixltx2e}
\usepackage{comment}
\usepackage{multirow, booktabs}
\usepackage{siunitx}
\usepackage{dblfloatfix}
\usepackage{xargs}  

\usepackage[colorinlistoftodos,prependcaption]{todonotes}
\newcommandx{\change}[2][1=]{\todo[linecolor=blue,backgroundcolor=blue!25,bordercolor=blue,#1]{#2}}

\usepackage{tikz}
\usetikzlibrary{matrix}
\usetikzlibrary{calc,fit}

\setlist{leftmargin=3.0mm}

\usepackage{tikz}
\usepackage{calc}

\begin{document}

\title{WinoCNN: Kernel Sharing Winograd Systolic Array for Efficient Convolutional Neural Network Acceleration on FPGAs\vspace{-10mm}}

\author{
\IEEEauthorblockN{
Xinheng Liu\IEEEauthorrefmark{1},
Yao Chen\IEEEauthorrefmark{2},
Cong Hao\IEEEauthorrefmark{3},
Ashutosh Dhar\IEEEauthorrefmark{1},
Deming Chen\IEEEauthorrefmark{1},\IEEEauthorrefmark{2}
}
\IEEEauthorblockA{
\IEEEauthorrefmark{1}University of Illinois at Urbana-Champaign, IL, USA,
\IEEEauthorrefmark{2}Advanced Digital Sciences Center, Singapore\\
\IEEEauthorrefmark{3}Georgia Institute of Technology, GA, USA
}
\IEEEauthorblockA{
Email: xliu79@illinois.edu, yao.chen@adsc-create.edu.sg, callie.hao@gatech.edu, \{adhar2, dchen\}@illinois.edu
}
\vspace{-10mm}
}



\maketitle

\begin{abstract}
The combination of Winograd's algorithm and systolic array architecture has demonstrated the capability of improving DSP efficiency in 
accelerating convolutional neural networks (CNNs) on FPGA platforms. 
However, handling arbitrary convolution kernel sizes in FPGA-based Winograd processing elements and supporting efficient data access remain underexplored.
In this work, we are the first to propose an optimized Winograd processing element (WinoPE), which can naturally support multiple convolution kernel sizes with the same amount of computing resources and maintains high runtime DSP efficiency.
Using the proposed WinoPE, we construct a highly efficient systolic array accelerator, termed WinoCNN.
We also propose a dedicated memory subsystem to optimize the data access.
Based on the accelerator architecture, we build accurate resource and performance modeling to explore optimal accelerator configurations under different resource constraints.
We implement our proposed accelerator on multiple FPGAs, which outperforms the state-of-the-art designs in terms of both throughput and DSP efficiency.
Our implementation achieves DSP efficiency
up to 1.33 GOPS/DSP and throughput up to 3.1 TOPS with the Xilinx ZCU102 FPGA. These are 29.1\% and 20.0\% better than the best solutions reported previously, respectively.
\end{abstract}


\section{Introduction}
Convolution neural networks (CNN) have been playing an essential role in solving practical
applications, and FPGAs have demonstrated their flexibility, efficiency, and reconfigurability as an ideal platform for CNN acceleration~\cite{mlfpgaiot, platformchoice, cong2019dac, chen2019clouddnn, x2017High}. 
Many previous works have proposed different algorithms and architectures to achieve high performance for CNN acceleration on FPGAs~\cite{cong2019dac, chen2019clouddnn, DNNBuilder, li2016high, x2017High, skynet}.
Since DSPs in FPGAs are usually the major computational resource, the run-time DSP efficiency,
defined as the average amount of effective convolution operations executed per DSP per second (GOPS/DSP), is crucial for FPGA design performance and is one of the most important factors to evaluate the design quality for FPGAs~\cite{cong2019dac}~\cite{x2017High}~\cite{tdla}~\cite{h264}.

Meanwhile, Winograd's minimal filtering algorithm has been widely adopted in CNN acceleration~\cite{doi:10.1137/1.9781611970364}. 
It trades multiplications with additions to save computational resources~\cite{doi:10.1137/1.9781611970364}. 
In FPGA, such a trade-off saves DSP resources from massive amount of multiplications in CNNs, and hence improves the concurrency and efficiency of acceleration. 
However, due to the inherent characteristics of the algorithm, existing Winograd convolution algorithms are usually specifically designed for a fixed convolution kernel size, e.g., $3\mathsf{x} 3$~\cite{8839351}~\cite{sparsewino}. 
When applied to other popular kernel sizes, i.e., $1 \mathsf{x} 1$ in light-weight CNNs, it becomes inefficient due to the overhead of kernel padding ~\cite{sparsewino}.
In addition, the tile-based data pattern required by the Winograd algorithm together with the concurrent processing requirement usually result in high data transmission overhead~\cite{sparsewino}. 

Systolic array-based accelerator architectures are considered compelling to deal with the massive amount of computations and communications required by CNNs~\cite{polysa, autosys, sparsewino}, delivering the state-of-the-art performance. 
However, the performance of the systolic array-based architecture largely relies on the efficiency of the processing elements (PEs) inside the array, the data transmission among PEs, as well as the data access from external memory, which are all non-trivial to optimize.

In this work, to address the aforementioned issues and improve system performance and DSP efficiency for Winograd based CNN acceleration, 
we make the following contributions:
\begin{itemize}

\item We design a novel Winograd-based processing element, WinoPE, using our generalized resource sharing mechanism that supports flexible convolution kernel sizes with high DSP efficiency.


\item Using the proposed WinoPEs, we construct a scalable systolic array-based accelerator WinoCNN, which supports flexible configurations with different parallelism levels honoring FPGA resource constraints.

\item We design a fine-grained and highly efficient memory control system that can deal with different memory access patterns and provide tile-based data to our WinoPEs with high efficiency and throughput.

\item We propose accurate models for resource and performance estimation, which guide the design space exploration for the configurable parameters of our WinoCNN accelerator. 

\end{itemize}

\section{Background and Design Challenges}\label{sec:back}


\subsection{Winograd Convolution on FPGA}
\begin{figure}[]
    \centering
    \includegraphics[width=0.4\textwidth]{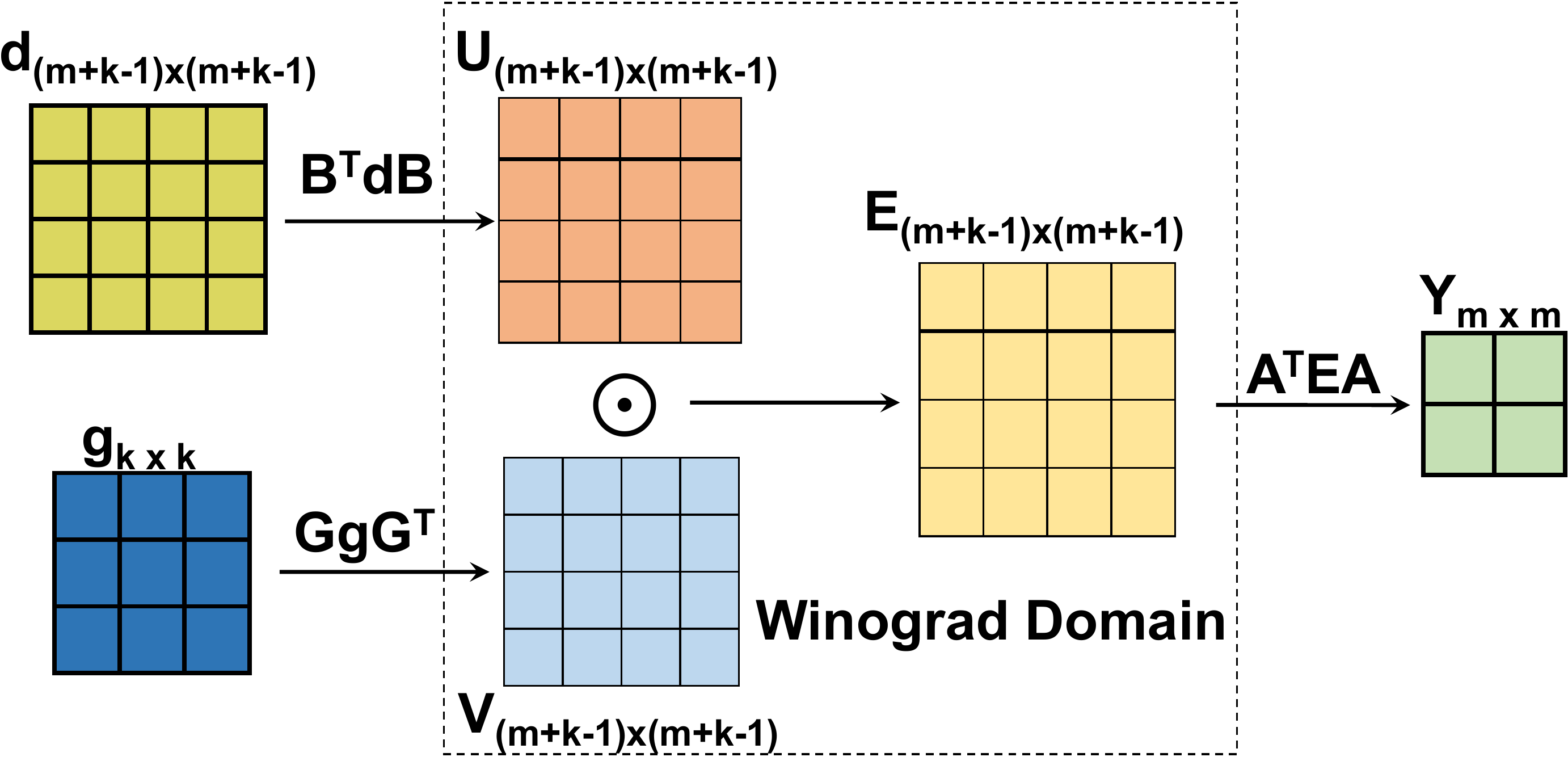}
    \caption{$F(m\mathsf{x} m,k\mathsf{x} k)$ Winograd convolution.}
    \label{fig:wino_eq}
    \vspace{-20pt}
\end{figure}

Winograd convolution is based on Winograd minimal filtering algorithm that 
computes an $m\mathsf{x} m$ output matrix $Y$ by convolving a $(m\text{+}k\text{-}1)\mathsf{x}(m\text{+}k\text{-}1)$ input matrix $d$ with a $k\mathsf{x} k$ kernel $g$ as described in Figure~\ref{fig:wino_eq}.
The input size is also treated as the Winograd filter size.
It reduces the number of multiplications at the cost of additions~\cite{doi:10.1137/1.9781611970364}.  
A 2D Winograd algorithm $F(m \mathsf{x} m, k \mathsf{x} k)$ includes a consecutive sequence of matrix transformation and element-wise multiplication (represented as $\odot$).
The $G$, $B$, and $A$ are constant transform matrices generated by Cook-Toom algorithm~\cite{doi:10.1137/1.9781611970364}. 


A convolution layer in CNN with $k\mathsf{x} k$ convolution kernel can be computed using Winograd algorithm with a configuration of $F(m\mathsf{x} m,k\mathsf{x} k)$.
The computation of each output feature-map $O$ with size $H_o \mathsf{x} W_o$ is divided into tiles with size $m\mathsf{x} m$, resulting in $\lceil{H_o/m}\rceil\lceil{W_o/m}\rceil$ tiles in each output channel.   
The computation of the output tile $O_{o,x_o,y_o}$ starting at pixel $(x_o,y_o)$ in channel $o$ can be completed by applying Winograd algorithm on input tiles $I_{i,x_i,y_i}$ starting at pixel $(x_i,y_i)$ in all the $C$ input feature-map channels with kernel $K_{o,i}$ and summing the results up, as shown in Eq.~\ref{eq:convlayer}.
\vspace{-3pt}
\begin{equation}\label{eq:convlayer}
\scalebox{0.8}{
     $ O_{o,x_o,y_o} =
     A^T(\sum_{i=1}^{C}[(B^T\cdot I_{i,x_i,y_i} \cdot B) \odot (G \cdot K_{o,i} \cdot G^T)])A $
}
\vspace{-3pt}
\end{equation}

The transformation operations with $B, G, A$ are matrix multiplications with constant element values that can be completed by add/shifting operations. 
So the total number of multiplications equals to the number of element-wise multiplications of $U$ and $V$, which is less than the required multiplications in the conventional convolution~\cite{DBLP:journals/corr/Lavin15b}. 
Since the {multiplications} on FPGAs are conducted by DSPs, reducing required {multiplications} in convolution helps to improve parallelism with a given number of DSPs and hence improves computation performance.

However, there is a critical problem:
The constant transformation matrices ($B$, $G$, $A$) for a given convolution kernel size have fixed patterns; this results in inefficient DSP utilization when using the hardware designed for one kernel size to a different kernel size, where it has to either split/pad the input data/kernels or to instantiate a new accelerator.
For example, to compute $1\mathsf{x}1$ convolution kernel with a Winograd-based PE designed for $3 \mathsf{x} 3$ kernel, we need to pad $1 \mathsf{x} 1$ convolution to  $3 \mathsf{x} 3$ with zeros, which can only achieve $\frac{1}{9}$ of the DSP efficiency of executing $3\mathsf{x}3$ convolution; or alternatively, instantiating a dedicated accelerators for $1 \mathsf{x} 1$ kernel only, which occupies additional resources.
Hence, designing a {Winograd-based} PE with flexible support for different kernel sizes while maintaining high DSP efficiency is essential but remains unexplored.

\subsection{Systolic Architecture}

A systolic array~\cite{kung1979systolic} is typically composed of many interconnected identical PEs, where the intermediate data is computed by PEs and passed to adjacent PEs.
Systolic array architectures are efficient for parallel computing  and is widely adopted by FPGA accelerators for matrix multiplications and convolutions~\cite{ x2017High}~\cite{aydonat2017opencl}.
One previous design~\cite{sparsewino} proposes a systolic array architecture with Winograd algorithm to accelerate sparse convolution, which achieves $5\mathsf{x}$ higher performance compared to the normal dense convolution accelerator.
Another work~\cite{8839351} proposes a systolic array architecture specifically designed for ResNet units.

In general, mapping the application to systolic array requires the data buffering in the PEs and the short PE-to-PE data transmission pattern.
CNNs are not naturally providing such buffering and connections patterns, which requires careful refinement of the orders of the operations and buffering of the data.


\subsection{Efficient Memory Access}

Inefficient data access of the PEs downgrades the overall performance~\cite{fp-dnn,x2017High,TCAD20}. 
To support efficient data access with limited off-chip memory bandwidth, the memory subsystem for the accelerator must be carefully designed for specific data re-arrangements and access patterns~\cite{8839351}, e.g., using multiple line-buffers \cite{TCAD20}.
However, 
it is difficult to create a universal design that would be compatible with different CNN layer configurations.
In addition, the systolic array of PEs requires the memory subsystem to provide concurrent off-chip memory access and on-chip data reuse to fully utilize the computational capacity of all PEs. Winograd algorithm further complicates the memory access requirements due to the varied planar data access patterns of the PEs.

\section{Design Principles}\label{sec:prin}

To resolve the challenges discussed in Section~\ref{sec:back}, we design our WinoCNN accelerator system with the following design principles.

\subsection{Sharing in Winograd Algorithm}
As discussed in Section~\ref{sec:back}, the low DSP efficiency of the Winograd algorithm for varying convolutional kernel sizes is caused by the constant transformation matrices.
The key solution is to provide flexible kernel size support within the Winograd convolution PE without reloading the transformation matrix and reorganizing the computation procedure.
For a Winograd convolution $F(m\mathsf{x} m,k\mathsf{x} k)$, the transformation matrices $B,A,G$ and the intermediate Winograd filter sizes are fixed, as shown in Figure \ref{fig:wino_eq}. The required number of element-wise multiplications 
equals to the size of $U$ and $V$, which is $(m\text{+}k\text{-}1)\mathsf{x} (m\text{+}k\text{-}1)$.

The input transformation matrix $B$ depends on the size of input tile $d$ for the input transformation ($U\text{=}B^TdB$). 
For a set of Winograd algorithm configurations with a Winograd filter size $\omega$, 
denoted as $F_\omega(m \mathsf{x} m, k \mathsf{x} k)$, where $\omega\text{=}m$+$k$-$1$ $(\omega \geqslant k)$. 
As long as $\omega$ values are the same, the computation patterns of input transformation and element-wise multiplication are exactly the same. 
Matrices $B_\omega^T(m\mathsf{x}m,k\mathsf{x}k)$ with same $\omega$ are identical. An example for $\omega=4$ is shown in Figure \ref{fig:wino_mat}.
Meanwhile, $U$ and $V$ are all $\omega \mathsf{x}\omega$ matrices.
Therefore, the hardware resource to process $U=B_\omega^TdB_\omega$ and $E=U \odot V$ can be shared among all $F_\omega(m\mathsf{x} m, k \mathsf{x} k)$.

The transformation matrices $G$ and $A$ will be different for different convolutional kernel sizes under the same $\omega$. 
We observe that there are a large amount of repeated values for the $G$ and $A$ matrices across different m and k values when $\omega$ is the same, and the different element(s) could be used as identifier(s) for different kernel sizes and output sizes. 
As shown in Figure~\ref{fig:wino_mat}, a single element $s$ could be used to identify $G_{4}(4\mathsf{x}4, 1\mathsf{x}1)$ and $G_{4}(2\mathsf{x}2, 3\mathsf{x}3)$.
\textbf{Also, this sharing property of the transformation matrix $A_\omega$ and $G_\omega$ can be generalized to larger Winograd filter size $\omega$ such as $F_8$ and $F_{10}$ for larger convolution kernel sizes such as $5\mathsf{x}5$ and $7\mathsf{x}7$ with multiple identifiers.}
As shown in Figure \ref{fig:wino_mat_F6}, the transformation matrices $G_6$ and $A_6$ with three identifiers $s_0,s_1$ and $s_2$ can be shared for the convolution kernel sizes $1\mathsf{x}1$, $3\mathsf{x}3$ and $5\mathsf{x}5$.
This provides us a unique opportunity to reuse the same computation resource (DSP) for different input kernel sizes using a unified PE for Winograd convolution. 
The design details of the PE and resource sharing are presented in Section~\ref{sect:wino_conv}.

\tikzset{%
  highlight1/.style={rectangle,rounded corners,color=red!,fill=red!15,draw,fill opacity=0.0,thick,inner sep=0pt}
}
\tikzset{%
  highlight2/.style={rectangle,rounded corners,color=green!,fill=green!15,draw,fill opacity=0.1,thick,inner sep=1pt}
}

\tikzset{%
  highlight1_0/.style={rectangle,rounded corners,color=red!,fill=red!15,draw,fill opacity=0.0,thick,inner sep=2pt}
}
\tikzset{%
  highlight2_0/.style={rectangle,rounded corners,color=green!,fill=green!15,draw,fill opacity=0.1,thick,inner sep=1pt}
}
\tikzset{%
  highlight3_0/.style={rectangle,rounded corners,color=blue!,fill=blue!15,draw,fill opacity=0.2,thick,inner sep=0pt}
}

\tikzset{%
  highlight3/.style={rectangle,rounded corners,color=blue!,fill=blue!15,draw,fill opacity=0.2,thick,inner sep=2pt}
}

\begin{figure}[]
\label{fig:wino_mat}
\centering
\begin{small}
\begin{tikzpicture}
    \matrix (m) [matrix of math nodes, left delimiter=\lbrack, right delimiter=\rbrack,
     row sep=0 mm, nodes={minimum width=8pt, minimum height=8 pt }] {
        1 & 0 & -1  &0 \\
        0 & 1 & 1   &0  \\
        0 & -1& 1   &0\\
        0 & -1& 0   &1\\
      }; 
      \node[below of = m, yshift=0pt](e){ $B^T_4(4\mathsf{x}4,1\mathsf{x}1)$};
     \node[below of = m, yshift=-10pt](e){ $=B^T_4(2\mathsf{x}2,3\mathsf{x}3)$};
\end{tikzpicture}
\begin{tikzpicture}
    \matrix (m) [matrix of math nodes, left delimiter=\lbrack, right delimiter=\rbrack,
     row sep=0 mm, nodes={minimum width=8pt, minimum height=8 pt }] {
      1 & 0 &  0  \\
      \frac{1}{2}&  \frac{1}{2} & \frac{1}{2}   \\
        \frac{1}{2} & -\frac{1}{2} & \frac{1}{2}  \\
        s & 0 & 1  \\
      }; 
         \node[highlight2, fit=(m-1-1.north west) (m-4-3.south east)] {};
            \node[highlight1, fit=(m-1-1.north west) (m-4-1.south east)] {};

      \node[below of = m, yshift=0pt](e){\textcolor{red}{$\square$} s=1: $G_4(4\mathsf{x}4,1\mathsf{x}1)$};
      \node[below of = m,yshift=-10pt]{\textcolor{green}{$\square$} s=0: $G_4(2\mathsf{x}2,3\mathsf{x}3)$};
\end{tikzpicture}
\begin{tikzpicture}
    \matrix (m) [matrix of math nodes, left delimiter=\lbrack, right delimiter=\rbrack,
     row sep=0 mm, nodes={minimum width=8pt, minimum height=8 pt }] {
      1 & 1 &  1 &0 \\
      0 &  1 & -1  &s \\
        0 & 1 & 1 &0 \\
        0 & 1 & -1 &1 \\
      }; 
            \node[highlight1_0, fit=(m-1-1.north west) (m-4-4.south east)] {};
    \node[highlight2_0, fit=(m-1-1.north west) (m-2-4.south east)] {};

      \node[below of = m, yshift=0pt](e){\textcolor{red}{$\square$} s=0: $A^T_4(4\mathsf{x}4,1\mathsf{x}1)$};
      \node[below of = m,yshift=-10pt]{\textcolor{green}{$\square$} s=-1: $A^T_4(2\mathsf{x}2,3\mathsf{x}3)$};
\end{tikzpicture}
\end{small}
\caption{Winograd transformation matrix for $F_4$.}
\label{fig:wino_mat}
\end{figure}

\begin{figure}[]
\centering
\begin{small}
\begin{tikzpicture}
    \matrix (m) [matrix of math nodes, left delimiter=\lbrack, right delimiter=\rbrack,
     row sep=0 mm, nodes={minimum width=8pt, minimum height=8 pt }] {
\frac{1}{4}  &   0  &   0   &  0   &  0  \\
\frac{-1}{6} & \frac{-1}{6} &  \frac{-1}{6} & \frac{-1}{6} & \frac{-1}{6}\\
\frac{-1}{6} &  \frac{1}{6} &  \frac{-1}{6} & \frac{1}{6}  & \frac{-1}{6} \\
\frac{1}{24}& \frac{1}{12} &  \frac{1}{6 } & \frac{1}{3}  & \frac{2}{3 } \\
\frac{1}{24} & \frac{-1}{12}&  \frac{1}{6}  & \frac{-1}{3} & \frac{2}{3 }  \\
 s_0   &   0  &   s_1   &  0   &  s_2  \\
      }; 
            \node[highlight3, fit=(m-1-1.north west) (m-6-5.south east)] {};
                  \node[highlight2, fit=(m-1-1.north west) (m-6-3.south east)] {};
        \node[highlight1, fit=(m-1-1.north west) (m-6-1.south east)] {};

      \node[below of = m, yshift=-18pt](e){\textcolor{red}{$\square$} $s_0s_1s_2=100$: $G_6(6\mathsf{x}6,1\mathsf{x}1)$};
      \node[below of = m,yshift=-28pt]{\textcolor{green}{$\square$} $s_0s_1s_2=010$: $G_6(4\mathsf{x}4,3\mathsf{x}3)$};
       \node[below of = m,yshift=-38pt]{\textcolor{blue}{$\square$} $s_0s_1s_2=001$: $G_6(2\mathsf{x}2,5\mathsf{x}5)$};
\end{tikzpicture}
\begin{tikzpicture}
    \matrix (m) [matrix of math nodes, left delimiter=\lbrack, right delimiter=\rbrack,
     row sep=0 mm, nodes={minimum width=8pt, minimum height=8 pt }] {
1  &1  &1   &1   & 1   &0\\
0  &1  &-1  &2   &-2   &s_0\\
0  &1  &1   &4   & 4   &0\\
0  &1  &-1  &8   &-8   &s_1\\
0  &1  &1   &16  &16   &0\\
0  &1  &-1  &32  &-32  &s_2\\
      }; 
      \node[highlight1_0, fit=(m-1-1.north west) (m-6-6.south east)] {};
      \node[highlight2_0, fit=(m-1-1.north west) (m-4-6.south east)] {};
      \node[highlight3_0, fit=(m-1-1.north west) (m-2-6.south east)] {};
      \node[below of = m, yshift=-18pt](e){\textcolor{red}{$\square$} $s_0s_1s_2=001$: $A^T_6(6\mathsf{x}6,1\mathsf{x}1)$};
     \node[below of = m,
     yshift=-28pt](e){\textcolor{green}{$\square$} $s_0s_1s_2=010$: $A^T_6(4\mathsf{x}4,3\mathsf{x}3)$};
  \node[below of = m,
     yshift=-38pt](e){\textcolor{blue}{$\square$} $s_0s_1s_2=100$: $A^T_6(2\mathsf{x}2,5\mathsf{x}5)$};
\end{tikzpicture}
\end{small}
\caption{Winograd transformation matrix for $F_6$.}
\label{fig:wino_mat_F6}
\vspace{-20pt}
\end{figure}

\vspace{-4pt}
\subsection{Task Mapping For PEs}

\begin{figure}[]
\begin{subfigure}{0.4\textwidth}
    \centering
    \vspace{-6pt}
    \includegraphics[width=\textwidth]{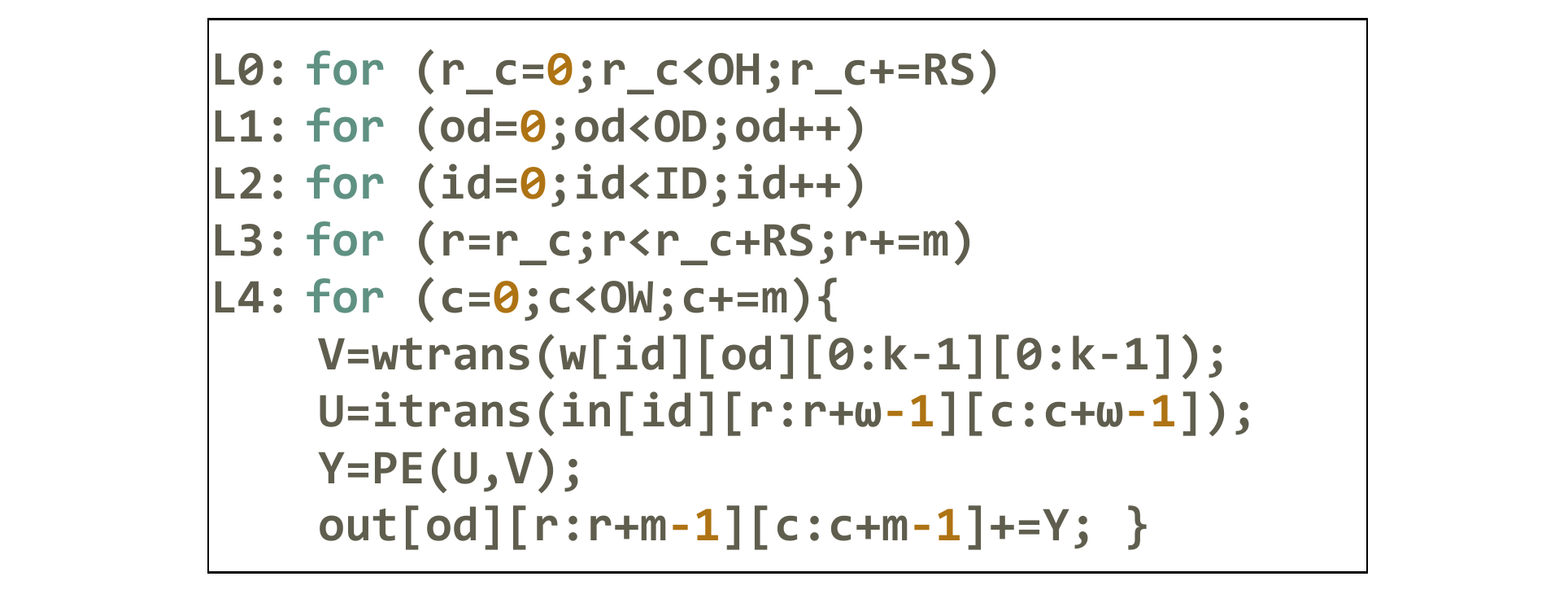}
    \caption{}
    \label{lst:conv_loop}
\end{subfigure}\\
\begin{subfigure}{0.4\textwidth}
    \centering
    \includegraphics[width=\textwidth]{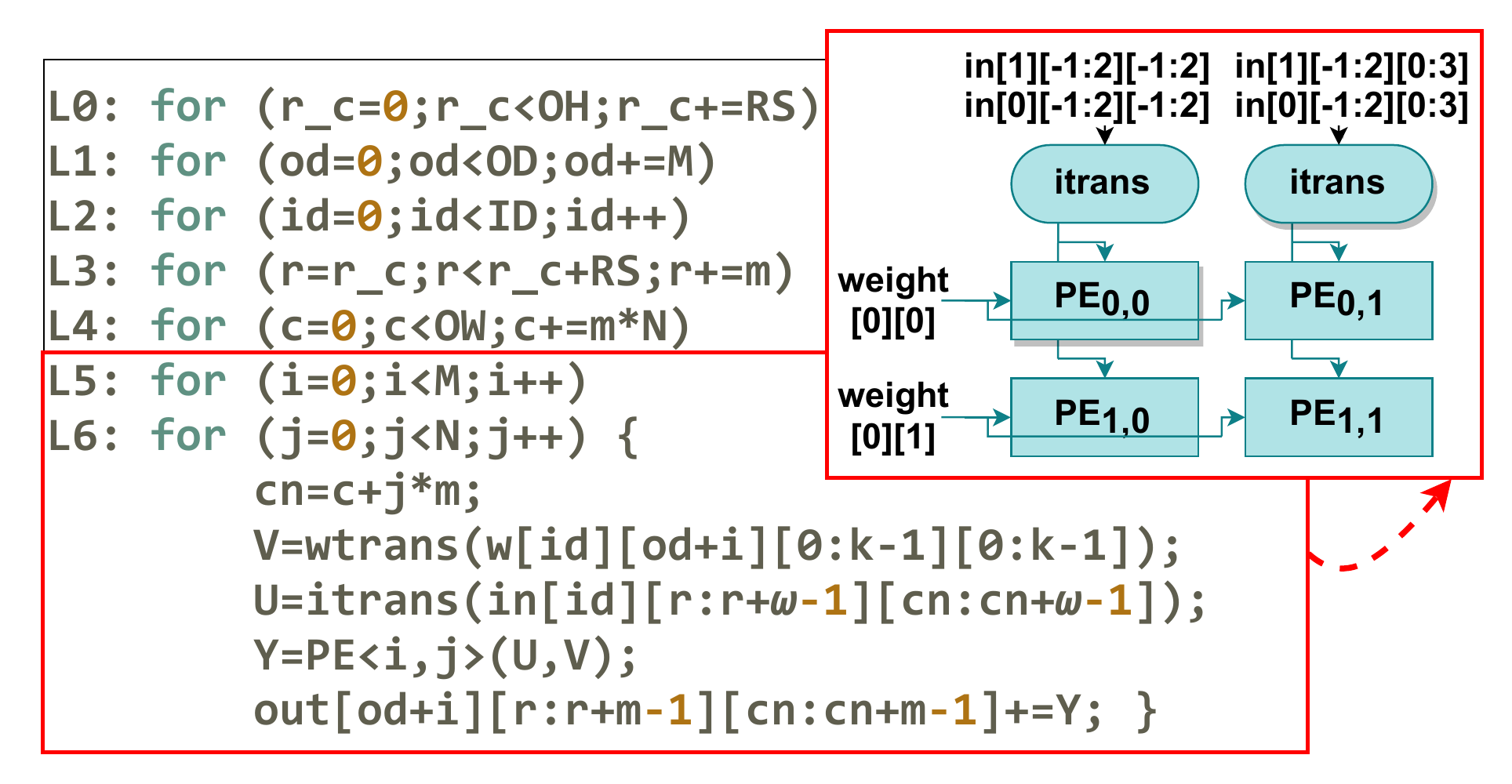}
    \caption{}
    \label{lst:win_loop}
\end{subfigure}
 \caption{Loops of computation process.}
    \vspace{-12pt}
\end{figure}

We assume a PE can perform the element-wise multiplication and output transformation $PE(U,V)=A^T(U\odot V)A$ for $F_\omega(m\mathsf{x} m, k \mathsf{x} k)$ Winograd convolution in one cycle.  
To properly map the convolution task into PEs, we partition the computation process of a convolution layer into several iterations. In each iteration, $RS$ consecutive rows of output feature map are computed.
Figure \ref{lst:conv_loop} shows the pseudo-code to compute the output feature map of a convolution layer with $k\mathsf{x}k$ kernel size using one PE.
The input, weight and output are represented as C-style array \texttt{in[ID][IH][IW]}, \texttt{w[ID][OD][k][k]}, and \texttt{out[OD][OH][OW]}, respectively.
However, loops shown in the Figure~\ref{lst:conv_loop} do not have the tiled structure to target the 2D PE array.
In order to map the computation to the 2D PE array and increase the parallelism of data processing, we rearrange the loop as shown in Figure \ref{lst:win_loop} and introduces two levels of tiling for the computation.
Loop \texttt{L0} iterates through the output rows with a step of $RS$.
Loop \texttt{L1} iterates through the output depth with a tile size of $M$.
Loop \texttt{L2} iterates through the input depth.
Loop \texttt{L3} segments the $RS$ output rows into {Winograd output tile} of size $m$.
Loop \texttt{L4} partitions the output columns into segments containing $N$ size-$m$ output tiles.
After unrolling of \texttt{L5} and  \texttt{L6}, $M\mathsf{x}N$ tiles of data will be processed by an $M\mathsf{x} N$ PE array in one cycle
(as shown in Figure~\ref{lst:win_loop} for a $2\mathsf{x}2$ array).
In this way, all WinoPEs with the same row index or column index share the same weights or the same input tile, respectively.

Note the direct mapping of the tiled computation to the PE array will generate high fanout (as shown in the embedded figure for the 2x2 array) and worsens
the timing of the implementation. 
In order to address this issue, we schedule the computation
of the PEs following the structure of an $M\mathsf{x}N$ systolic array.
The detailed design will be presented in the Section~\ref{sect:over_arch}.

\subsection{Efficient Memory Access}

\begin{figure}[]
    \centering
    \includegraphics[width=0.4\textwidth]{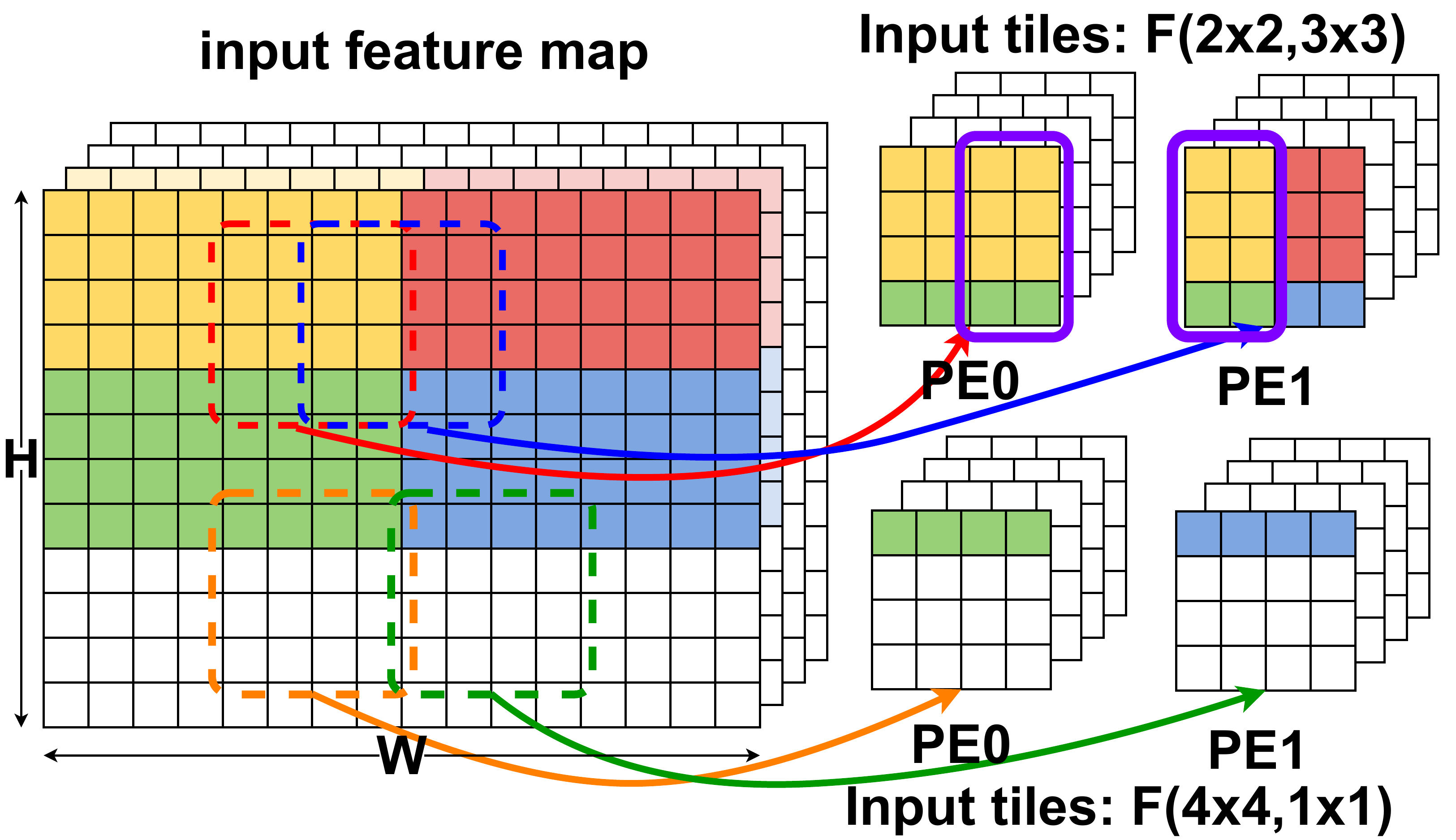}
    \caption{Planar data access pattern of PEs.}
    \label{fig:fold_feature}
    \vspace{-20pt}
\end{figure}

Efficient execution of Winograd-based PEs requires simultaneous data access within a tile, as shown in Figure ~\ref{fig:fold_feature}.
This planar data access pattern (data tiles) brings in a challenge for efficient memory control and data supply for the PEs.
When multiple PEs are instantiated as an array to process different tiles of a CNN layer, there are also overlaps among the data tiles required by the PEs.
As shown in the example in Figure~\ref{fig:fold_feature},
the data tiles required by adjacent PEs overlap with each other (marked with purple circle). Simply assigning input buffers for all the PEs would cause high on-chip memory usage~\cite{autosys}.
However, line buffer based design~\cite{TCAD20} faces difficulties when supplying multiple tiles for Winograd-based PEs under systolic architecture that requires varied memory access patterns, i.e., varied window moving steps.
As shown in Figure \ref{lst:win_loop} \texttt{L4}, the PE array requires $N$ input tiles with a horizontal moving step size of $N\cdot m$ each cycle.
Meanwhile, the output tile size $m$ differs according to kernel size $k$, leading to a varied window moving step. 
These motivate us to design a specialized memory system for our WinoCNN architecture. 

To design an efficient PE array to work with such data access patterns, we draw three design principles:
First, to improve computation efficiency with high parallelism, the data elements inside one tile must be fetched in parallel and provided to the computational unit simultaneously;
and second, the overlapped data across tiles shall be fetched from external memory only once and then reused to reduce memory access overhead.
Third, the memory system should be able to supply data for Winograd convolutions with different kernel sizes efficiently.
\section{Implementation}\label{sec:impl}


We implement our WinoPE, systolic array and memory subsystem based on the design principles from Section~\ref{sec:prin} to build our WinoCNN acceleration system.

\subsection{WinoPE: PE With Multiple Kernel Support} \label{sect:wino_conv}

\begin{figure}[]
    \centering
    \includegraphics[width=0.4\textwidth]{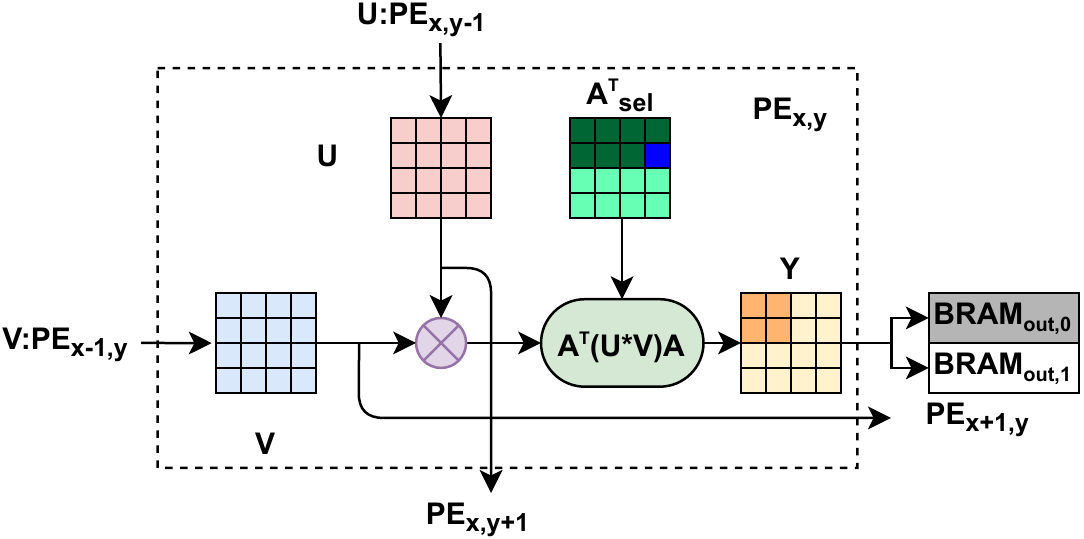}
    \caption{WinoPE processing logic.}
    \label{fig:wino_cell}
    \vspace{-24pt}
\end{figure}

Our WinoPE is the basic processing unit of the Winograd systolic convolution accelerator system (WinoCNN), where each WinoPE is able to complete the computation of an input kernel and a set of feature tiles in a single clock cycle.
WinoPE is featured with the flexible support for different convolution kernel sizes without the DSP overhead in a unified architecture.
As discussed in Section~\ref{sec:prin}, the Winograd algorithm with the same Winograd filter size $\omega$ can share the corresponding transformation matrices as well as the expensive dot product module.
We choose the design of sharing between $F_4(4\mathsf{x} 4, 1\mathsf{x} 1)$ and $F_4(2\mathsf{x} 2, 3\mathsf{x} 3)$ as the example to present our kernel sharing mechanism.
Figure \ref{fig:wino_cell} shows the unified architecture to process a single tile in our WinoPE.
It contains an input tile register array (red block), a weight tile register array (blue block), a matrix of multipliers (purple circle), an output transformation module (green block), and an output tile (dark and light yellow) towards an output buffer.

In each working cycle, the WinoPE reads in a tile of input data and a tile of weights in parallel.
Note here, the input tiles are transformed on-chip when they are fetched from input buffer and the convolution kernel weights are transformed before they are stored into the on-chip memory
to reduce the resource usage for transformation logic.
After fetching the input and weight, the element-wise multiplication $U\odot V$ is performed.
The output transformation module takes the results of $U \odot V$ and generates the output tile.
The data fetching and element-wise multiplication modules can be directly shared and fully utilized by different convolution kernel sizes.
To handle the different output caused by different kernel sizes, we design a selectable output transformation matrix $A^T_{sel}$, in which the \textbf{selection bit} $s$ in the matrix $A$ is used as a matrix identifier, as shown in Figure~\ref{fig:wino_mat} (Section~\ref{sec:prin}).
As an instance in $F_4$, when $s$ is set to 0, the WinoPE performs $F_4(4\mathsf{x} 4,1\mathsf{x} 1)$ algorithm, where the whole $4\mathsf{x} 4$ matrix is the output of the WinoPE (light yellow block in Figure~\ref{fig:wino_cell}). When $s$ is set to -1, the WinoPE performs $F_4(2\mathsf{x} 2,3\mathsf{x} 3)$ algorithm, where the top left four elements of the result matrix is the output (dark yellow block).
In this way, our WinoPE processes convolution layers with different kernel sizes without DSP overhead.
Finally, the computed outputs are stored in the output buffer constructed with BRAMs.
Note that such a selection bit design
can be easily extended to the Winograd algorithm with larger Winograd filter sizes for larger convolution kernel sizes.

Furthermore, we partition the input tile register $U$ and weight tile register $V$ into individual registers that each contains a single data from the input channels, so that the multiplication for an entire tile is finished in a single clock cycle. 
The processing efficiency of WinoPE is further increased by instantiating $Q$ input channels of batch size $B$ of tile registers with the corresponding number of $U\odot V$ multiplier matrices.
An adder tree constructed with LUT is used to accumulate the outputs from the multiplier matrices.

The selection bit design allows us to share the computation resources without wasting the DSPs when processing convolutions with different kernel sizes. 
However, it remains challenging to use Winograd convolution algorithm for large kernel convolution and irregular kernel convolution. 
A practical limitation is the larger Winograd filter size requires more LUT resources to conduct the addition operation during the constant matrix multiplication for Winograd convolution algorithm.
Also, recent DNN models adopt irregular convolution kernels such as $1\mathsf{x}7$ and $7\mathsf{x}1$ sizes that are not well supported by the Winograd algorithm.

To handle the large kernel convolution and irregular kernel convolution, we design a split mechanism that splits the target convolution kernel into supported kernel sizes as shown in the Equation \ref{eq:split_kernel} and \ref{eq:split_conv}. \vspace{-6pt} 
\begin{equation}
\label{eq:split_kernel}
\vspace{-6pt}
\begin{tiny}
\begin{gathered}
K_s^{i,j}[h][w]=\begin{cases}
        0, \begin{tabular}{c}
            $ik\text{+}h \geq H_t $ or 
            $jk\text{+}w \geq W_t$
        \end{tabular}\\
         K_t[ik\text{+}h][jk\text{+}w],  otherwise 
    \end{cases}
\end{gathered}
\end{tiny}
\end{equation}

\begin{equation}
\label{eq:split_conv}
\begin{tiny}
\begin{gathered}
Output_{target}= \sum_{i,j} FM^{ik,jk} \ast K_s^{i,j}
\end{gathered}
\end{tiny}
\vspace{-4pt}
\end{equation}
$K_t$ represents the target convolution kernel with size $H_t\mathsf{x} W_t$ and $K_s$ represents the supported convolution kernel with size $k \mathsf{x} k$.
The target kernel is split into $\lceil{\frac{H_t}{k}}\rceil\mathsf{x}\lceil{\frac{W_t}{k}}\rceil $ supported kernels with unaligned elements padded with zeros. 
The split kernel $K_s^{i,j}$ is segmented from the target kernel with $ik,jk$ offset from the top left element. The targeted convolution (denoted as $\ast$) is completed by applying convolution for each supported kernel $K_s^{i,j}$ on input features with the same 2D pixel offset (denoted as $FM^{ik,jk}$) and summing up the split results as shown in Equation \ref{eq:split_conv}.

\begin{figure}[]
    \centering
    \vspace{-10pt}
    \includegraphics[width=0.4\textwidth]{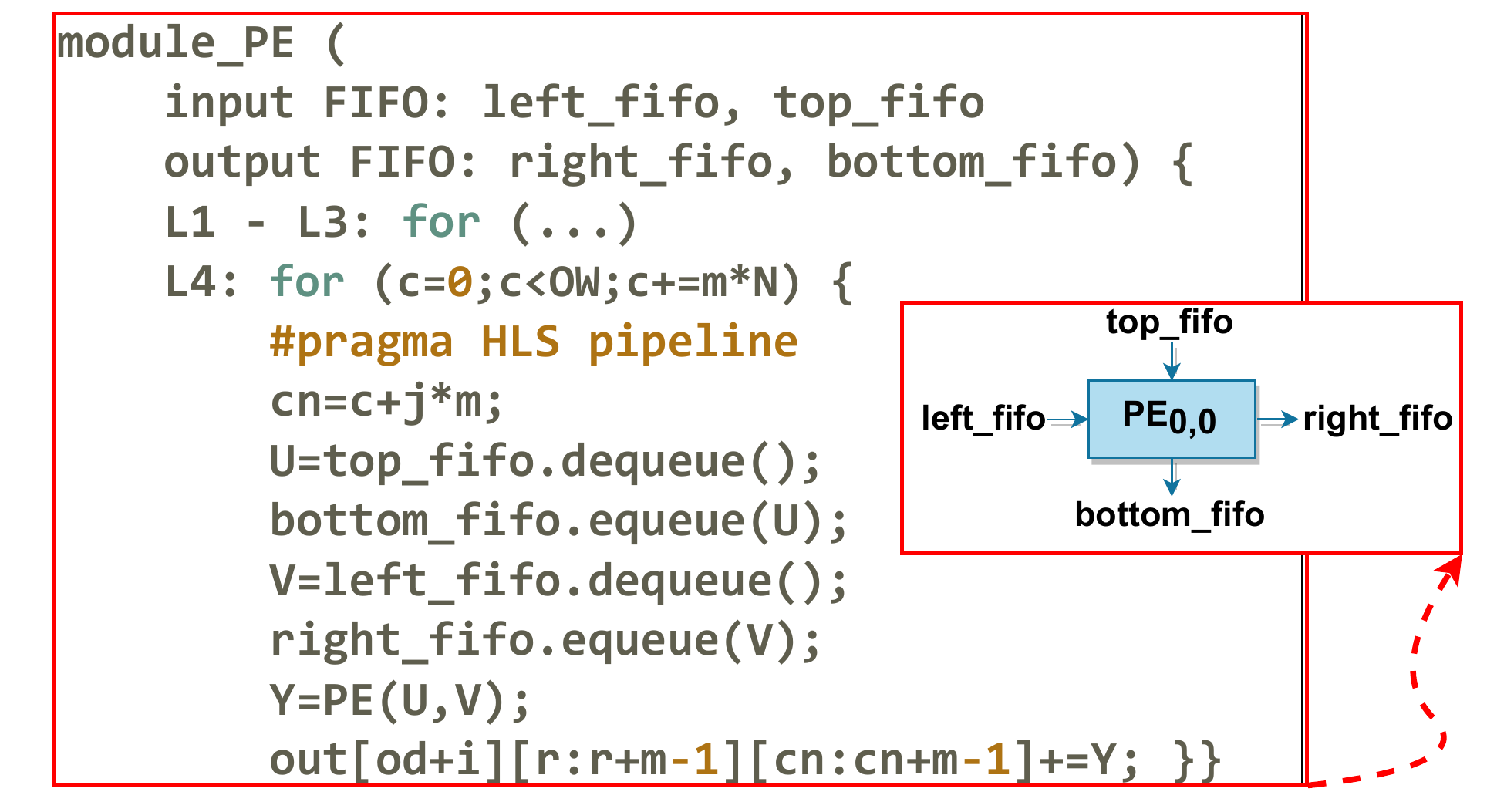}
       \vspace{-6pt}
    \caption{Loop behaviour for PE$_{i,j}$}
    \label{lst:pe_loop}
    \vspace{-20pt}
\end{figure}

\subsection{Parameterized Systolic Array} \label{sect:over_arch}

Instead of sharing a single set of data tiles among different WinoPEs in the same clock cycle (as shown in the Figure~\ref{lst:win_loop}), we construct the WinoPEs as an $M\mathsf{x} N$ systolic array that shares the weight and input data among WinoPEs by shifting them PE-to-PE
to further utilize the on-chip registers and reduce the high fanout and long connection caused by the flattened implementation.
To achieve this, we take advantage of the insensitivity of the loop order and assign FIFOs to WinoPEs (as shown in Figure~\ref{lst:pe_loop}).
Note here, the PEs are called by an outside loop as \texttt{L0} and the loops \texttt{L1-L4} are the same as the ones in Figure \ref{lst:win_loop}; however, the input Winograd tile and the weight tile are fetched from the top and left FIFO interfaces which connect to the top and left neighbours of the WinoPEs, respectively.
In the same iteration, the input and weight data are pushed into bottom and right FIFO interfaces and passed to the bottom and right neighbours after one clock cycle  
due to the blocking mechanism of the FIFO. 
Therefore, the WinoPEs are constructed as a systolic array.


With the assigned FIFOs for our WinoPEs, we could easily instantiate the systolic WinoPE array by organizing the row and column FIFOs, denoted as $row\_fifo[M][N]$ and $col\_fifo[N][M]$. 
The $M, N$ parameters are configurable during the WinoCNN system generation.

\subsection{Hierarchical Memory Subsystem} \label{sect:data_opt}

 
To provide the required data to the WinoPE array efficiently,
we propose:
(1) a BRAM buffer matrix that has a unique addressing mechanism to support efficient parallel data access, and (2) a pipelined planar data control and scheduling to provide efficient on-chip data reuse and support the flexible input tile access pattern.


\subsubsection{BRAM Buffer Matrix}

\begin{figure}[]
    \centering
    \vspace{-6pt}
    \includegraphics[width=0.48\textwidth]{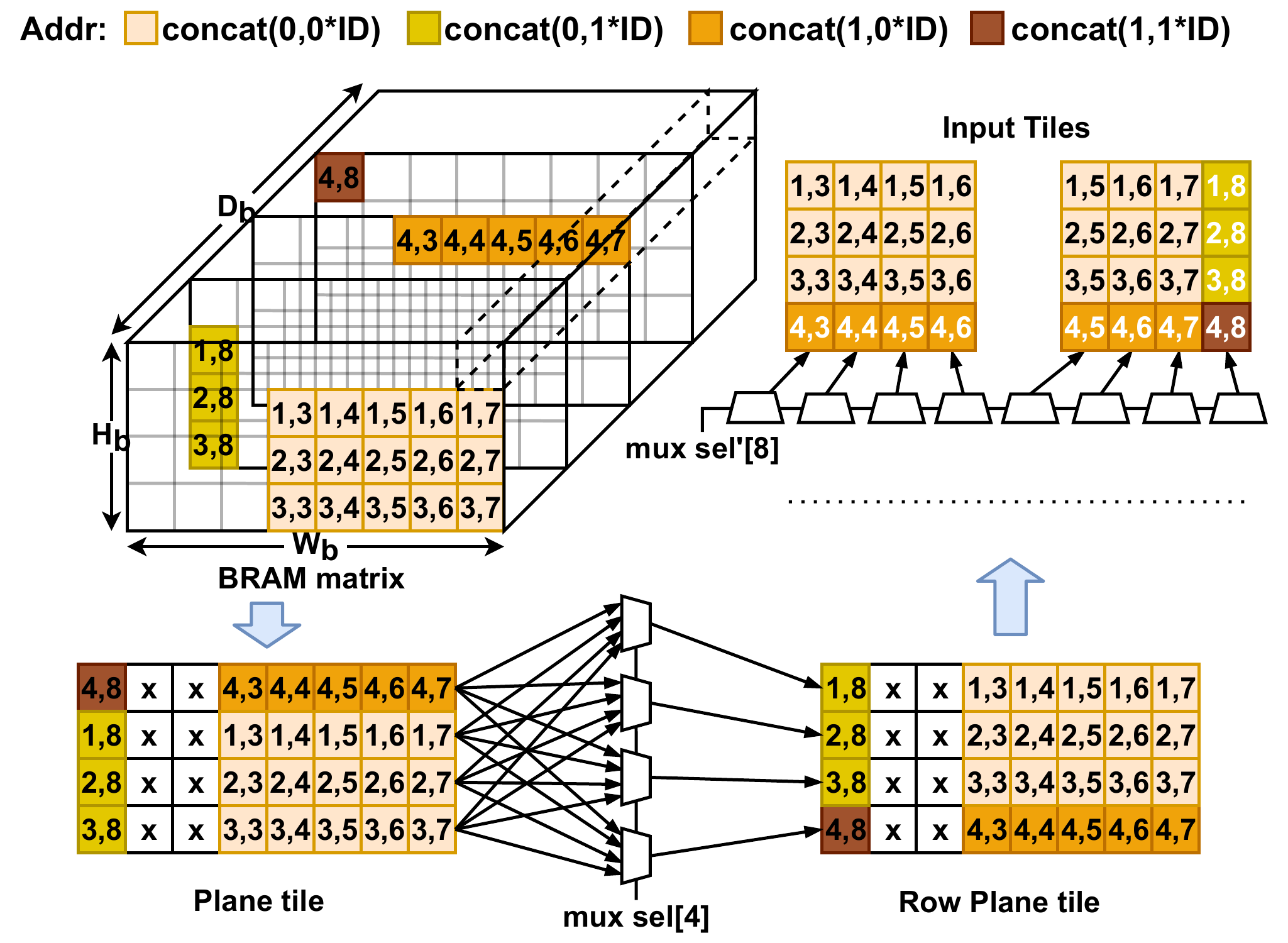}
    \caption{Hierarchical Memory Subsystem.}
    \label{fig:loc_map}
    \vspace{-18pt}
\end{figure}

To guarantee the parallel access of the input tiles, we fold the input feature-maps into a matrix of BRAM buffers, 
denoted as $BRAM_{in}[H_b][W_b][D_b]$, which consists of $H_b \mathsf{x} W_b$ BRAM buffer instances (or BRAM bank) of depth $D_b$, as shown in Figure~\ref{fig:loc_map}.
Each BRAM buffer instance has its individual address port and data port, hence total $H_b \mathsf{x} W_b$ entries can be accessed from the BRAM buffer matrix in each cycle at different addresses.
An address mapping mechanism is designed as shown in Eq.~\ref{eq:bram_map} to decide the location in the buffer as $BRAM_{in}[h][w][addr]$ for a certain input pixel in the feature map $in[id][r][c]$, where $ID,id,r,c$ represent the number of input channel for a layer, channel index, row index and column index for the pixel in the input feature map :
\begin{equation}\label{eq:bram_map}
\begin{tiny}
\begin{gathered}
h=r \% H_b, ~
w=c \% W_b\\
addr = concat(\left \lfloor{r/H_b}  \right \rfloor, \left \lfloor{c/W_b} \right\rfloor\cdot ID+id )\\ 
\end{gathered}
\end{tiny}
\end{equation}
The BRAM banks in the same row share the same high address bits, while the BRAM banks in the same column share the same low address bits.
The concatenation function ensures that the entries are accurately located with the given index.
With the sequence of input Winograd tiles denoted as $T_N$, where $T_N$ consists of $N$ tiles with $\omega\mathsf{x}\omega$ size, defined in Eq. \ref{eq:in_tile},
all the elements in $T_N$ forms a continuous input data block, denoted as $T_U$, with size $\omega\mathsf{x} ((N\text{-}1)m\text{+}\omega)$:

\begin{equation}
\label{eq:in_tile}
\begin{tiny}
\begin{gathered}
T_N[n]\text{=}
in[id][r\text{:}r\text{+}\omega\text{-}1][c\text{+}nm\text{:}c\text{+}nm\text{+}\omega\text{-}1]
\end{gathered}
\end{tiny}
\end{equation}
\begin{equation} \label{eq:inplane_blk}
\begin{tiny}
\begin{gathered}
T_U\text{=}\bigcup_{n\text{=}0}^{N-1}T_N[n]\text{=}
in[id][r\text{:}r\text{+}\omega\text{-}1][c\text{:}c\text{+} (N\text{-}1)m\text{+}\omega\text{-}1] 
\end{gathered}
\end{tiny}
\end{equation}

As an instance in Figure~\ref{fig:loc_map}, $H_b\text{=}4$, $W_b\text{=}8$, $\omega\text{=}4$, $m\text{=}2$ and $N\text{=}2$ and two data tiles are required to be accessed within input feature maps $in\_data[0][1\text{:}4][3\text{:}6]$ and $in\_data[0][1\text{:}4][5\text{:}8]$. 
The union of the two input tiles can be represented as $T_U\text{=}in\_data[0][1\text{:}4][3\text{:}8]$. 
According to the address mapping defined in Eq.~\ref{eq:bram_map}, the pixels in $in\_data[0][1\text{:}4][3\text{:}8]$ are accessed from four different regions of BRAM buffer matrix in one clock cycle with different high address bits and low address bits. 



\subsubsection{Planar Data Access}
The input tiles are then passed through a 3-stage pipeline to ensure the data reuse and to provide the planar data to the WinoPEs.
As shown in Figure \ref{fig:loc_map},
The first stage stores the $H_b \mathsf{x} W_b$ output from BRAM matrix buffer into registers. 
The second stage ensures the row order of the planar data with a row plane multiplexer array.
The third stage splits the plane to tiles by a column multiplexer array.
The mux selection bits are generated on-the-fly regarding the values of $r$, $c$, and $id$. Since both the BRAM bank addresses and the mux selection are generated on-the-fly, the memory architecture is able to supply input tiles with varied window moving steps regarding the kernel size for the current convolution layer.

\section{System architecture and modeling}\label{sec:inti_model}

We construct our WinoCNN system and build
the performance and resource models for easy exploration of the architectural configurations.

\subsection{WinoCNN Architecture Overview} \label{sec:winocnn_system}
\begin{figure}[]
    \centering
    \includegraphics[width=0.35\textwidth]{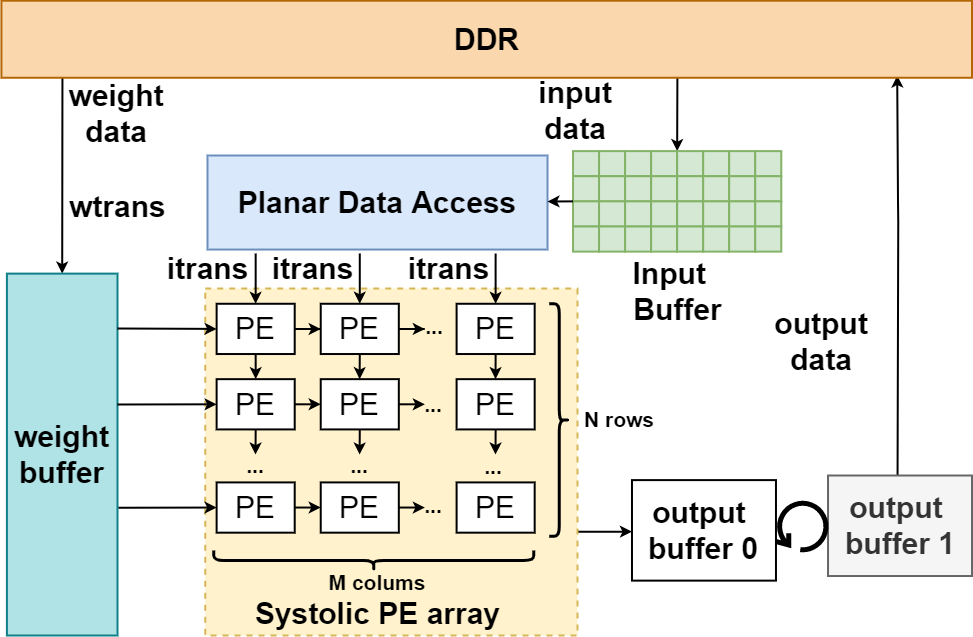}
    \caption{Overall architecture.}
    \label{fig:architect}
    \vspace{-18pt}
\end{figure}

The overall architecture of our WinoCNN accelerator system is shown in Figure \ref{fig:architect}.
Note here, the flexible convolution kernel size support is provided by our WinoPEs.
The convolution layers of the input models are computed in output row stationary. 
The input reading, computation, and output data offloading are scheduled to run in parallel.

\subsection{System Modeling}\label{subsec:modeling}

As shown in Section~\ref{sec:impl}, our system is built with performance and resource sensitive architectural parameters, the corresponding models are built for design space exploration.

\subsubsection{Resource Model}

\noindent\textbf{DSP usage.}
The major instances of DSPs are occupied by the WinoPEs. 
Each WinoPE computes the element-wise sum of the product along $Q$ input channels for input tiles $U^{\omega \mathsf{x} \omega}$ of batch size $B$ and weight tiles $V^{\omega \mathsf{x} \omega}$.
Thus, the total number of DSPs required by a WinoPE is $\omega^2 \cdot B \cdot Q $.
The systolic WinoPE array contains $M \mathsf{x} N$ WinoPEs, so the total DSPs required by our WinoCNN accelerator is:
\vspace{-6pt}
\begin{equation}
\begin{tiny}
\begin{gathered}
    DSP_{use}= \omega^2\cdot M\cdot N\cdot B\cdot Q    
\end{gathered}
\end{tiny}
\end{equation}
\noindent\textbf{BRAM occupation.}
The BRAM resource is mainly occupied by the input, weight and output buffers. 
The BRAMs are in the form of 18-bit width and 1024 depth blocks.

The input buffer is a buffer matrix of size $H_b \mathsf{x} W_b$ with buffer depth as $D_{in}$. Each bank should be capable of storing $B$ input data with 8 bits. The number of BRAMs for input is $H_b\cdot W_b\cdot \lceil{8\cdot B/18}\rceil \cdot \lceil D_{in} /1024\rceil $.

Each row of the systolic array requires $\omega^2\cdot Q$ transformed and quantized 16-bit weight data, thus requires $\lceil{\omega^2\cdot Q\cdot 16/18}\rceil$ BRAM blocks with a fixed depth of 1024 to provide enough weight access bandwidth. 
The total BRAM required for weight buffer is 
$M \cdot \lceil{16 \cdot \omega^2 \cdot Q /18}\rceil$.

Each WinoPE has a $\omega\mathsf{x}\omega$ buffer matrix to store 18-bit temporary output data of batch $B$ and buffer depth $D_{out}$. 
With the requirement of 2 buffers for ping-pong access, each WinoPE needs $2\cdot \lceil{\omega^2\cdot B\cdot 18/18}\rceil\cdot \lceil D_{out} /1024\rceil$ BRAMs as output buffer.

The total number of BRAM is the sum of the above:
\begin{equation}
\begin{tiny}
\begin{gathered}
    BRAM_{use}= H_bW_b\lceil{8\cdot B/18}\rceil \cdot \lceil D_{in} /1024\rceil \\
    + M\lceil{ 16\cdot\omega^2Q/18}\rceil+2\cdot MN\omega^2B\lceil D_{out} /1024\rceil
\end{gathered}
\end{tiny}
\end{equation}

\subsubsection{Latency Model}
Communication latency $t_{comm}$ and computation latency $t_{comp}$ in each phase of the convolution procedure are used to build the latency model.
The maximum value between these two in each phase dominates the overall latency. 

Since all weights are required once within each loop iteration, we have $D_{weight}=k^2\cdot ID\cdot OD$. 
Each loop iteration includes a read input process and a write output process with the corresponding data transmission amount of $D_{input}=RS\cdot ID \cdot IW \cdot B$ and $D_{output}=RS\cdot OD \cdot OW \cdot B$.
Neglecting the absence of output writing  in the first iteration and the input  reading in the last iteration, we estimate the communication latency as:
\begin{equation}
\begin{tiny}
\begin{gathered}
     t_{comm}=(D_{weight}+D_{input}+D_{output})/BW   
\end{gathered}
\end{tiny}
\end{equation}
To compute $RS$ rows of outputs in each computation process, the WinoPE array needs to sum up the convolution results through $ID$ input planes to generate $RS\mathsf{x} OW$ output data for $OD$ output planes. 
In each cycle, $M\mathsf{x} N$ WinoPEs sum up the convolution results of $Q$ input planes for $N\mathsf{x} m\mathsf{x} m$ output pixels along $M$ depth. 
Considering implementation frequency $f$, the computation latency for each iteration is:
\begin{equation}
\begin{tiny}
\begin{gathered}
      t_{comp}= \lceil{ID/Q}\rceil  
    \lceil{OD/M}\rceil
    \left\lceil{RS/m}\right \rceil
    \left\lceil{OW/(N\cdot m)}\right \rceil/f   
\end{gathered}
\end{tiny}
\end{equation}
The overall latency is estimated as:
\begin{equation}
\begin{tiny}
\begin{gathered}
t_{loop} = \lceil{OH/RS} \rceil \cdot max(t_{comm}, t_{comp})
\end{gathered}
\end{tiny}
\end{equation}
\subsubsection{Parameter Exploration}

For the convenience of hardware implementation, we fix the batch size at $B=2$. 
To guarantee the access of the planar data, we set $H_b$ as 4 or 8 for $F_4\text{ or }F_6$ respectively and $W_b=\min 2^k, \text{ s.t. } W_b \geq 2\omega$.
%
%
Note here, the row step $RS$ is a variable during the processing and is chosen as large as possible so that the input and output rows can fully utilize the on-chip buffers.
For a given CNN model, the $M,N,Q,D_{in}$ and $D_{out}$ are explored targeting $min({\sum}_{l\in layers} t_{loop,l})$ with the given DSP and BRAM resources on the platform.

\section{Evaluations} \label{sect:exp_and_rst}

To validate the effectiveness of our design, we use Xilinx ZCU102 and Ultra96 boards for evaluation, where both platforms are equipped with a quad-core ARM Cortex-A53. The detailed resource specifications are shown in Table~\ref{tab:config_perf}.
We use Vivado HLS design suit 2019.2 for accelerator implementation using C++.

\subsection{WinoPE Evaluation}

\subsubsection{Resource effectiveness}
We first compare the resource utilization of our WinoPE with the PEs without multiple kernel support, as shown in Table~\ref{tab:singlepecom}. All PEs are configured with $Q$=4 and $B$=2. 
The same DSP utilization in each PE type ensures that the maximum parallelism of the PEs is the same. 
Using the same amount of DSP resources, our WinoPE 
consumes more LUT and FF resources than each dedicated PE but with the benefit of supporting different convolution kernel sizes without effecting the runtime efficiency.


\setlength{\tabcolsep}{4pt}
\begin{table}[h]
    \centering
    \vspace{-6pt}
    \caption{Resource utilization of different PEs}
    \begin{small}
    \begin{tabular}{c|c|c|c||c|c|c|c}
        \hline
        \hline
        PE type         &   LUT     & FF       & DSP & PE type         &   LUT     & FF       & DSP   \\ \hline
        $F_4(2\mathsf{x}2,3\mathsf{x}3)$      &   5328&   2430         & 128   &$F_6(4\mathsf{x}4,3\mathsf{x}3)$      &   21542  &   19235    & 288    \\\hline
        $F_4(4\mathsf{x}4,1\mathsf{x}1)$      &   6495  &  9831          & 128  & $F_6(6\mathsf{x}6,1\mathsf{x}1)$      & 24056  &    39126      & 288 \\\hline
        WinoPE-$F_4$          &  7852     &    10501     & 128 & WinoPE-$F_6$    &  33959    &    42793    & 288   \\\hline
        \hline
    \end{tabular}
    \end{small}
    \label{tab:singlepecom}
    \vspace{-6pt}
\end{table}



\setlength{\tabcolsep}{5pt}
\begin{table*}[]
\centering
\begin{small}
\caption{WinoCNN configuration and performance on different platforms}
\label{tab:config_perf}
\vspace{-6pt}
\begin{tabular}{c|c|c|c|c|c|c|c|c|c|c|c|c|c}
\hline
\hline
\multirow{2}{*}{Platform} & \multicolumn{5}{c|}{PE Config.} & \multicolumn{5}{c|}{Resource Util.(\% of (total))} & \multicolumn{3}{c}{Throughput (GOPS) } \\ \cline{2-14} 
& M & N & Q &  $D_{in}$ & $D_{out}$ & DSP& BRAM & LUT  & FF & Freq.(MHz) & VGG-16 & INet-V4 & YoloV2   \\ \hline
Ultra96(WinoPE-$F_4$) &  2  &   1  &  4 &  4096  &  1024  & 77.8(360) & 85.9(432)    &  60.8(70K)  &      43.2(141K) & 250  &  265   &127.2  &157.5 \\ \hline
ZCU102(WinoPE-$F_4$)& 8 & 2 & 4 &  8192 & 1024& 82.8(2520)&   95.5(1824) & 76.3(274K) & 43.4(548K)  & 250 & 1862  & 820.3  & 1241\\ \hline
ZCU102(WinoPE-$F_6$) & 4 & 2 & 4 & 4096 &1024 & 93(2520) & 87(1824) & 81(274K)  & 48(548K)  & 214  & 3120.3  & 857.23  & 1717.7 \\ \hline
\hline
\end{tabular}%
\end{small}
\vspace{-6pt}
\end{table*}

\begin{table*}[]
    \centering
\begin{small}

    \setlength{\tabcolsep}{1mm}{
    \caption{Comparison with state-of-the-art designs}
    \vspace{-6pt}
    \label{tab:perf_alex}
    \begin{tabular}{c|c|c|c|c|c|c|c|c|c|c|c} 
        \hline 
        \hline
        & \cite{TCAD20} &  \cite{autosys} &  \cite{8839351} &  \cite{sparsewino} & \cite{tvlsi20} &\multicolumn{3}{c|}{Vitis-AI~\cite{vitisai}}  & \multicolumn{3}{c}{Ours. (WinoPE-F\textsubscript{6})} \\ \hline
        Platform & ZCU102 & Arria10 GT1150 & Stratix V GSMD8 & XCVU095 & Arria-10& \multicolumn{3}{c|}{ZCU102} &  \multicolumn{3}{c}{ZCU102}  \\\hline 
        Model & VGG-16 & VGG-16 & Resnet-18 & VGG-16 & VGG-16 & VGG-16 &INet-V4& YoloV2 & VGG-16  & INet-V4 & YoloV2 \\\hline 
        \begin{tabular}{c} Freq. (MHz) \end{tabular}& 200 & 231.85 & 160 & 150 & 250& \multicolumn{3}{c|}{281} & \multicolumn{3}{c}{214}  \\\hline 
        Precision& 16-bit fixed&  8-16 bit fixed & 16-bit fixed  &8-16-bit fixed & 16-bit fixed &\multicolumn{3}{c|}{8 bit}  & \multicolumn{3}{c}{8-16 bit} \\\hline
        Batch size& 32 & -  &  -   &-  & -& \multicolumn{3}{c|}{2} &  \multicolumn{3}{c}{2} \\ \hline
        DSP Usage & 2520 & 1500 &  576   &768 &1344 &\multicolumn{3}{c|}{1926} & \multicolumn{3}{c}{2345} \\ \hline 
        \hline
        Thro. (GOPS) & 2601.3$^{1}$ & 1171.3   & 233   &460 & 1642  &1225.2  & 1390 & 1008 & \textbf{3120.3} & 857.23 & \textbf{1717.7}  \\\hline
        
        Latency (ms) &10.43$^{1}$ &26.85 &7.23 &- &- & 57.53 & 35.26 & 16  &\textbf{19.67}  & 49.7 & \textbf{13.9}  \\ \hline
        \begin{tabular}{c} DSP Eff. (GOPS/DSP) \end{tabular} & 1.03$^{1}$ & 0.780 & 0.405 & 0.599 &1.22 &0.636 &0.722 & 0.523 & \textbf{1.33} & 0.388 & \textbf{0.73}\\ \hline
        \hline
    \end{tabular}}\\
    \footnotesize{$^{1}$The throughput, latency and DSP efficiency are only for the convolutional layers.}
    \end{small}  
    \vspace{-16pt}
\end{table*}


\subsubsection{Performance effectiveness} 
\begin{figure}[h]
\vspace{-6pt}
\centering
\pgfplotsset{width=0.48\textwidth,height= 0.25\textwidth, compat=1.17,
tick label style={font=\small},
label style={font=\small},
legend style={font=\tiny},
}
\begin{tikzpicture}
\begin{axis}[
ybar,
bar width=1.5pt,
enlargelimits=0.15,
legend image code/.code={
\draw [#1] (0cm,-0.1cm) rectangle (0.1cm,0.1cm);
},
legend style={at={(0.04,0.98)},
draw=none,
legend style={nodes={scale=0.65, transform shape}},
anchor=north west},
ylabel={DSP efficiency (GOPS/DSP) },
ylabel near ticks,
legend cell align={left},
ylabel style={font=\small},
symbolic x coords={1x1,1x3,3x3,5x5,1x7,7x7,9x9},
xtick=data,
xlabel={Convolution kernel sizes},
]
\addplot coordinates {(1x1,0.2) (1x3,0.2) (3x3,0.2) (5x5,0.2)(1x7,0.2)(7x7,0.2)(9x9,0.2)};
\addplot coordinates {(1x1,0.2) (1x3,0.2) (3x3,0.2) (5x5,0.2)(1x7,0.2)(7x7,0.2)(9x9,0.2)};
\addplot coordinates {(1x1,0.05) (1x3,0.15) (3x3,0.45) (5x5,0.3125)(1x7,0.2)(7x7,0.2722)(9x9,0.45)};
\addplot coordinates {(1x1,0.089) (1x3,0.2666) (3x3,0.8) (5x5,0.555)(1x7,0.2071)(7x7,0.483)(9x9,0.8)};
\addplot coordinates {(1x1,0.179) (1x3,0.180) (3x3,0.4383) (5x5,0.3090)(1x7,0.2)(7x7,0.269)(9x9,0.446)};
\addplot coordinates {(1x1,0.179)(1x3,0.2634) (3x3,0.8) (5x5,0.555)(1x7,0.205)(7x7,0.483)(9x9,0.8)};
\legend{ \text{F\textsubscript{4}(4x4,1x1)},
\text{F\textsubscript{6}(6x6,1x1)},\text{F\textsubscript{4}(2x2,3x3)},\text{F\textsubscript{6}(4x4,3x3)},WinoPE-F\textsubscript{4},WinoPE-F\textsubscript{6},
}
\end{axis}
\end{tikzpicture}
\vspace{-6pt}
\caption{DSP efficiency of WinoPE to theoretical values.}
\label{fig:comp_eff}
\end{figure}
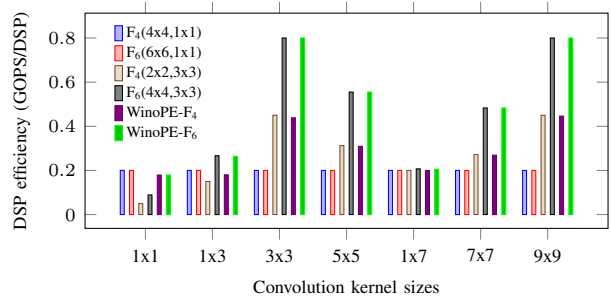

Since the DSP efficiency without efficient data supply will be lower than the theoretical performance, we evaluate our WinoPE together with our memory subsystem and compare the results to other PEs theoretical performance with the assumption of the data supply is perfect.
We first conduct experiments of synthetic convolution layers with different kernel sizes and compare them to the theoretical performance values using the same configuration as shown in Figure~\ref{fig:comp_eff}.
We measure the DSP efficiency to exclude the impact of different platforms with a system frequency at 100Mhz.
The DSP efficiency of WinoPEs for different kernel sizes is measured on board, and the maximum performance for other PEs are calculated theoretically; both are shown in Figure~\ref{fig:comp_eff}.
Compared to the theoretical performance of $F_4$ and $F_6$, our implementations of WinoPE-$F_4$ and WinoPE-$F_6$ under all kernel sizes achieve near-maximal theoretical performance with the proposed memory subsystem. 

\vspace{-8pt}
\subsection{WinoCNN Evaluation}

We adopt the most representative CNN models as benchmarks to demonstrate the effectiveness of our WinoCNN system design, including VGG-16, Inception-V4 (denoted as INet-V4), and YoloV2.
The non-convolution layers are executed in the processors with multi-thread optimization for end-to-end model execution. 
All convolutional layers are executed on the WinoCNN accelerator.

\subsubsection{WinoCNN configuration}
We explore the optimal WinoCNN system configurations for different platforms using our analytical model.
The selected values are shown in Table~\ref{tab:config_perf} together with resource utilization and runtime performance on the different platforms for different models.

The WinoCNN accelerator configurations for different platforms and different Winograd kernel sizes vary significantly because of the different DSP and BRAM capacity of the platforms, where
all configurations target to fully utilize the on-chip DSP and BRAM resources.
The achievable frequency under each configuration for a certain platform is also shown together with the final performance.
Our WinoCNN system naturally supports better timing due to the timing-friendly shorter data path between the WinoPEs.
Notably, for the networks with homogeneous convolutional layers, i.e., VGG-16, our design achieves 3.12 TOPS throughput at 214MHz clock frequency while the performance drops to 857.23 GOPS when there are multiple divergent convolutional layer configurations in Inception-V4, i.e., $1\mathsf{x}7$ kernel.
This is because of the varied efficiency of the Winograd algorithm for different convolution kernel sizes.


\subsubsection{Comparison with state-of-the-art designs}
We then measure the execution latency, throughput, and DSP efficiency of our implemented models and compare them with the state-of-the-art implementations, as shown in Table \ref{tab:perf_alex}.

Since all the convolution layers are executed by our WinoCNN accelerator, the DSP efficiency and latency data are calculated for the convolution layers.
DSP efficiency of our design is $1.71\mathsf{x}$ of the design in~\cite{autosys}, which does not use Winograd transformation.
When compared to the designs with Winograd algorithm~\cite{sparsewino}~\cite{TCAD20} together with additional model-specific optimizations, our design 
shows a $1.2\mathsf{x}$ and a $6.78\mathsf{x}$ improvement of throughput compared to that of \cite{TCAD20} and \cite{sparsewino}, respectively.
Notably, the design in \cite{TCAD20} adopts a 32 batch size for FC layers, which is much larger than ours (fixed at 2) and leads to a long latency for a single image to be processed completely.
In the comparisons, all the previous architectures containing model-specific designs can not support flexible kernel sizes, while our WinoCNN supports multiple convolution kernels without effecting the DSP efficiency.
Our design also provides slightly better achievable frequency due to the efficient systolic array architecture on Xilinx platforms.

When comparing to the Vitis-AI implementations~\cite{vitisai}, our WinoCNN shows better throughput and latency for both VGG-16 and YoloV2 even with a lower clock frequency and without DSP double pumping.
For the Inception-V4 model which contains unique kernel shapes, i.e., $1\mathsf{x}7$, $7\mathsf{x}1$, $3\mathsf{x}1$ and $1 \mathsf{x} 3$, we use the less efficient $F(4\mathsf{x} 4, 1\mathsf{x} 1)$ or $F(6\mathsf{x} 6, 1\mathsf{x} 1)$ to process them, which lead to a worse performance than the specially optimized Vitis-AI processing cores.
\section{conclusion}\label{sec:conclusion}
\vspace{-2pt}
In this work, we present a systolic array based convolution accelerator design targeting the Winograd algorithm. 
Our accelerator, WinoCNN, is constructed by unique Winograd convolution PEs (WinoPE) which support flexible convolution kernel sizes without sacrificing DSP efficiency. WinoCNN also has an efficient memory subsystem that is suitable for planar data access for the array of WinoPEs.
Our accelerator system is configurable for different FPGA platforms with accurate resource and performance models. 
Overall, our accelerator delivers high throughput and state-of-the-art DSP efficiency comparing with previous accelerator implementations.
Our code release can be found at https://github.com/xliu0709/WinoCNN.

\section{acknowledgement}
\vspace{-4pt}
This work is supported in part by the IBM-Illinois Cen- \\
ter for Cognitive Computing Systems Research (C3SR), Semiconductor Research Corporation (SRC) and is also partially supported by the National Research Foundation, Prime Minister's Office, Singapore under its Campus for Research Excellence and Technological Enterprise (CREATE) programme.

\setstretch{0.75}
\newcommand{\BIBdecl}{\setlength{\itemsep}{0.25em}}
\bibliographystyle{IEEEtran}
\bibliography{IEEEabrv,ref}

\end{document}